\PassOptionsToPackage{unicode}{hyperref}
\PassOptionsToPackage{hyphens}{url}
\PassOptionsToPackage{dvipsnames,svgnames,x11names}{xcolor}
\documentclass[
  english,
  ,man,floatsintext]{apa6}
\usepackage{xcolor}
\usepackage{amsmath,amssymb}
\usepackage{iftex}
\ifPDFTeX
  \usepackage[T1]{fontenc}
  \usepackage[utf8]{inputenc}
  \usepackage{textcomp} 
\else 
  \usepackage{unicode-math} 
  \defaultfontfeatures{Scale=MatchLowercase}
  \defaultfontfeatures[\rmfamily]{Ligatures=TeX,Scale=1}
\fi
\usepackage{lmodern}
\ifPDFTeX\else
\fi
\IfFileExists{upquote.sty}{\usepackage{upquote}}{}
\IfFileExists{microtype.sty}{
  \usepackage[]{microtype}
  \UseMicrotypeSet[protrusion]{basicmath} 
}{}
\makeatletter
\@ifundefined{KOMAClassName}{
  \IfFileExists{parskip.sty}{%
    \usepackage{parskip}
  }{
    \setlength{\parindent}{0pt}
    \setlength{\parskip}{6pt plus 2pt minus 1pt}}
}{
  \KOMAoptions{parskip=half}}
\makeatother
\makeatletter
\ifx\paragraph\undefined\else
  \let\oldparagraph\paragraph
  \renewcommand{\paragraph}{
    \@ifstar
      \xxxParagraphStar
      \xxxParagraphNoStar
  }
  \newcommand{\xxxParagraphStar}[1]{\oldparagraph*{#1}\mbox{}}
  \newcommand{\xxxParagraphNoStar}[1]{\oldparagraph{#1}\mbox{}}
\fi
\ifx\subparagraph\undefined\else
  \let\oldsubparagraph\subparagraph
  \renewcommand{\subparagraph}{
    \@ifstar
      \xxxSubParagraphStar
      \xxxSubParagraphNoStar
  }
  \newcommand{\xxxSubParagraphStar}[1]{\oldsubparagraph*{#1}\mbox{}}
  \newcommand{\xxxSubParagraphNoStar}[1]{\oldsubparagraph{#1}\mbox{}}
\fi
\makeatother
\usepackage{graphicx}
\makeatletter
\newsavebox\pandoc@box
\newcommand*\pandocbounded[1]{
  \sbox\pandoc@box{#1}%
  \Gscale@div\@tempa{\textheight}{\dimexpr\ht\pandoc@box+\dp\pandoc@box\relax}%
  \Gscale@div\@tempb{\linewidth}{\wd\pandoc@box}%
  \ifdim\@tempb\p@<\@tempa\p@\let\@tempa\@tempb\fi
  \ifdim\@tempa\p@<\p@\scalebox{\@tempa}{\usebox\pandoc@box}%
  \else\usebox{\pandoc@box}%
  \fi%
}
\def\fps@figure{htbp}
\makeatother
\NewDocumentCommand\citeproctext{}{}

\makeatletter
 \let\@cite@ofmt\@firstofone
 \def\@biblabel#1{}
 \def\@cite#1#2{{#1\if@tempswa , #2\fi}}
\makeatother
\newlength{\cslhangindent}
\newlength{\csllabelwidth}
\newenvironment{CSLReferences}[2] 
 {\begin{list}{}{%
  \setlength{\itemindent}{0pt}
  \setlength{\leftmargin}{0pt}
  \setlength{\parsep}{0pt}
  \ifodd #1
   \setlength{\leftmargin}{\cslhangindent}
   \setlength{\itemindent}{-1\cslhangindent}
  \fi
  \setlength{\itemsep}{#2\baselineskip}}}
 {\end{list}}
\usepackage{calc}

\ifLuaTeX
\usepackage[bidi=basic,shorthands=off]{babel}
\else
\usepackage[bidi=default,shorthands=off]{babel}
\fi
\ifLuaTeX
  \usepackage{selnolig} 
\fi
\providecommand{\tightlist}{%
  \setlength{\itemsep}{0pt}\setlength{\parskip}{0pt}}
\usepackage{upgreek}
\usepackage{longtable}
\usepackage{lscape}
\usepackage{multirow}		
\usepackage{tabularx}		
\usepackage[flushleft]{threeparttable}	
\usepackage{threeparttablex}            

\makeatletter
\newcommand\LastLTentrywidth{1em}
\newlength\longtablewidth
\newcommand{\getlongtablewidth}{\begingroup \ifcsname LT@\roman{LT@tables}\endcsname \global\longtablewidth=0pt \renewcommand{\LT@entry}[2]{\global\advance\longtablewidth by ##2\relax\gdef\LastLTentrywidth{##2}}\@nameuse{LT@\roman{LT@tables}} \fi \endgroup}

\makeatletter
\renewcommand{\paragraph}{\@startsection{paragraph}{4}{\parindent}%
  {0\baselineskip \@plus 0.2ex \@minus 0.2ex}%
  {-1em}%
  {\normalfont\normalsize\bfseries\itshape\typesectitle}}

\renewcommand{\subparagraph}[1]{\@startsection{subparagraph}{5}{1em}%
  {0\baselineskip \@plus 0.2ex \@minus 0.2ex}%
  {-\z@\relax}%
  {\normalfont\normalsize\itshape\hspace{\parindent}{#1}\textit{\addperi}}{\relax}}
\makeatother

\makeatletter
\usepackage{etoolbox}
\patchcmd{\maketitle}
  {\section{\normalfont\normalsize\abstractname}}
  {\section*{\normalfont\normalsize\abstractname}}
  {}{\typeout{Failed to patch abstract.}}
\patchcmd{\maketitle}
  {\section{\protect\normalfont{\@title}}}
  {\section*{\protect\normalfont{\@title}}}
  {}{\typeout{Failed to patch title.}}
\makeatother

\usepackage{xpatch}
\makeatletter
\xapptocmd\appendix
  {\xapptocmd\section
    {\addcontentsline{toc}{section}{\appendixname\ifoneappendix\else~\theappendix\fi: #1}}
    {}{\InnerPatchFailed}%
  }
{}{\PatchFailed}
\makeatother
\usepackage{lineno}

\usepackage{csquotes}
\usepackage{amsmath}
\usepackage[utf8]{inputenc}
\usepackage[T1]{fontenc}
\usepackage[backend=biber,useprefix=true,style=apa,url=false,doi=false,sorting=nyt,eprint=false]{biblatex}
\DeclareLanguageMapping{american}{american-apa}
\usepackage{fancyvrb}
\usepackage{newfloat}
\usepackage{graphicx}
\usepackage{caption}
\usepackage{listings}
\usepackage{glossaries}
\makeglossaries
\DeclareFloatingEnvironment[fileext=los, listname=List of Schemes, name=Listing, placement=!htbp, within=section]{listing}
\usepackage{placeins}
\usepackage{todonotes}
\usepackage{bookmark}
\IfFileExists{xurl.sty}{\usepackage{xurl}}{} 
\hypersetup{
  pdftitle={How accurate are Bayes factor-based null hypothesis tests? A simulation study},
  pdfauthor={Daniel J. Schad1 \& Martin Modrák2},
  pdflang={en-EN},
  pdfkeywords={Bayes factors, Bayesian model comparison, simulation-based calibration, null hypothesis testing, crossed random effects for subjects and items},
  colorlinks=true,
  linkcolor={Maroon},
  filecolor={Maroon},
  citecolor={Blue},
  urlcolor={blue},
  pdfcreator={LaTeX via pandoc}}

\title{How accurate are Bayes factor-based null hypothesis tests? A simulation study}
\author{Daniel J. Schad\textsuperscript{1} \& Martin Modrák\textsuperscript{2}}
\date{}

\shorttitle{Bayes factor accuracy}

\affiliation{\vspace{0.5cm}\textsuperscript{1} Institute for Mind, Brain and Behavior, HMU Health and Medical University, Potsdam, Germany\\\textsuperscript{2} Second Faculty of Medicine, Charles University, Prague, Czech Republic}

\note{

Author Note \newline Correspondence concerning this article should be addressed to Daniel J. Schad, HMU Health and Medical University, Potsdam, Germany. E-mail: \href{mailto:danieljschad@gmail.com}{\nolinkurl{danieljschad@gmail.com}}. A preprint of this manuscript has been published on arXiv (\href{https://arxiv.org/abs/2406.08022}{https://arxiv.org/abs/2406.08022}). Reproducible code is available from \url{https://osf.io/3g86r/}. This study was not preregistered.

}

\abstract{%
Bayes factor null hypothesis tests provide a viable alternative to frequentist measures of evidence quantification.
Bayes factors for realistic data sets in areas like psychology cannot be calculated exactly and require numerical approximations to complex integrals. Crucially, the accuracy of these approximations, i.e., whether an approximate Bayes factor corresponds to the exact Bayes factor, is unknown, and may depend on data, prior, and likelihood. We have recently developed a novel statistical procedure, namely marginal simulation-based calibration (SBC) for Bayes factors,
to test whether the computed Bayes factors for a given analysis are accurate.
Here, we use marginal SBC for Bayes factors and calibration plots to test for some common cognitive designs, whether Bayes factors are calculated accurately. We use the bridgesampling/brms packages in R. We run analyses for three commonly used designs in psychology and psycholinguistics: (a) a design with random effects for subjects only, (b) a Latin square design with crossed random effects for subjects and items, but a single fixed-factor, and (c) a Latin square 2x2 design with crossed random effects for subjects and items. We find that Bayes factor estimates turn out accurate in cases when the bridgesampling algorithm does not issue a warning message, but can be biased and variable when a warning message is shown. These results support the use of brms/bridgesampling for null hypothesis Bayes factor tests in commonly used factorial designs. They also suggest that when a warning message is issued, Bayes factor results should not be trusted. The results show that it is practical to check whether Bayes factors are computed correctly.
}

\begin{document}
\maketitle

\section{Introduction}\label{introduction}

Bayes factors have emerged as a popular tool for quantifying evidence for hypotheses and comparing models in statistical data analysis (Chow \& Hoijtink, 2017; Hoijtink \& Chow, 2017; Lee, 2011; Mulder \& Wagenmakers, 2016; Vandekerckhove, Rouder, \& Kruschke, 2018). Bayes factors offer several advantages, such as their ability to assess the probability of the null hypothesis, incorporate prior information, and provide continuous measures of evidence strength. As a result, Bayes factors are increasingly used in many scientific domains, including psychology, neuroscience, and social sciences ({Heck et al.}, 2023).

However, computing exact Bayes factors is often not feasible, as it involves evaluating complex integrals that cannot be solved analytically, and approximations are necessary. For some restricted families of models and prior distributions, Bayes factors can be converted to relatively well-behaved low-dimensional integrals that are then directly evaluated numerically --- an example of this approach is the BayesFactor (Morey \& Rouder, 2022) R package. For more general models, we typically need to first estimate posterior distributions for both models with a suitable Markov-chain Monte Carlo (MCMC) sampler and then use the posterior samples to build an approximation to the marginal likelihood in a separate computational step --- an example of this approach is the bridgesampling (Gronau, Singmann, \& Wagenmakers, 2020) R package. In the former case, the precision of the approximation is limited by the precision of numerical integration, in the latter case the approximation is affected by imprecisions in both the MCMC step and in the marginal likelihood computation. Nevertheless, the quality of these approximations are as yet unclear and remain a concern, as biased approximations could lead to erroneous conclusions about the evidence for hypotheses and the relative strengths of models (Schad, Nicenboim, Bürkner, Betancourt, \& Vasishth, 2023).

To investigate the accuracy of Bayes factor estimates, Schad et al. (2023) recently proposed a method called marginal Simulation-Based Calibration (marginal SBC) for Bayes factors (for an alternative see Sekulovski, Marsman, \& Wagenmakers, 2024). Marginal SBC provides a means to assess whether Bayes factors are estimated accurately, or whether there is liberal or conservative bias in the Bayes factor estimator. In essence, marginal SBC works by defining a model prior (i.e., a priori model probabilities) and a parameter prior for each model (i.e., prior densities for each model parameter), and then doing a simulation many times, where in each simulation, one performs the following steps: (1) sample a model from the model prior, (2) for the sampled model, sample parameters from their prior distributions, (3) use the sampled parameters to simulate data from the model, (4) compute the Bayes factor using the method to be evaluated, and then use these Bayes factor estimates to compute the posterior model probabilities for each of the models given the simulated data. We can then compare the average posterior probabilities (averaged across simulations) with the prior probability; this allows marginal SBC to detect whether the Bayes factor estimates are biased or not. If the average posterior probability is equal to the prior probability, then this is consistent with accurate Bayes factor estimation. If the average posterior deviates from the prior, then this indicates a liberal or conservative bias in the Bayes factor estimation.

An important application domain of Bayes factor estimation is in the analysis of factorial designs using linear mixed-effects models (LMMs, Baayen, Davidson, \& Bates, 2008). LMMs are now increasingly used in experimental psychology and related fields to study the outcome from experiments with factorial designs, and have developed into a standard approach, which often replaces previous ANOVA-based analyses.
Nested model comparisons for balanced factorial designs can be analyzed using Bayes factors, which can replace traditional p-values as a means of assessing main effects, interactions, and specific contrasts (Schad, Vasishth, Hohenstein, \& Kliegl, 2020). In this context, Bayes factors are often used for null hypothesis tests, where a full model containing all fixed effects of interest is compared to a null model that lacks some effect or a set of effects ({van Doorn, Aust, Haaf, Stefan, \& Wagenmakers}, 2021).

In this paper, we aim to investigate whether Bayes factor estimates obtained from common factorial designs are biased or accurate using marginal SBC as well as calibration plots that can be derived from it (Modrák, Stroppel, \& Bürkner, 2025). We build upon our prior work (Schad, Nicenboim, \& Vasishth, 2024), which showed that data aggregation can lead to biased inferences in Bayesian linear mixed models and ANOVA. To further assess the accuracy of Bayes factor estimation, we focus on a pair of widely used R packages: bridgesampling (Gronau et al., 2020) in combination with brms (Bürkner, 2017). Specifically, we apply marginal SBC to a range of experimental repeated measures designs, including a 2x2 design with random effects for subjects, a design with a single fixed factor and crossed random effects for subjects and items, and a 2x2 design with crossed random effects for subjects and items.

We focus on these designs here because they are the dominant type of experimental design in cognitive psychology and linguistics. The dominance of such designs and statistical tools in these areas is clear from the fact that tutorial papers like Pinheiro and Bates (2006), Baayen et al. (2008), and Bates, Mächler, Bolker, and Walker (2015) continue to be widely referenced in major journals like
\emph{Cognition} and the \emph{Journal of Memory and Language} and in classic textbooks such as Gelman and Hill (2007) and Gelman et al. (2013). That linear mixed models continue to be a very important tool is clear from the steady stream of articles relating this approach that keep appearing in major psychology and psycholinguistic journals. A recent example illustrating the interest in mixed models is the special issue exclusively devoted to this topic in the official journal of the Society for Mathematical Psychology, Computational Brain and Behavior; see {van Doorn et al.} (2021).

Returning to our main goal in the present paper, by using marginal SBC, we can determine whether the Bayes factor estimates for the fixed effects are unbiased and provide a solid basis for statistical data analysis in these experimental designs. This investigation is important, as biased Bayes factor estimates can lead to erroneous conclusions about the evidence for hypotheses and the relative strengths of models. Overall, our study contributes to the growing literature on the application of Bayesian methods in the cognitive sciences (psychology, linguistics, and related areas).

To foreshadow our results: we found that Bayes factor estimates for linear mixed models based on the R-packages bridgesampling/brms were accurate when no warning message was issued by the bridge sampler. However, when a warning message was issued, Bayes factor estimates could turn out biased \textcolor{black}{due to a highly variable computation of the log marginal likelihoods}.
These results suggest that the accuracy of Bayes factor estimates in commonly used software tools cannot be taken for granted and that warning messages should be taken seriously. More generally, we suggest (cf. Schad et al., 2023) that marginal SBC for Bayes factors should be employed to test the accuracy for a given set of design, models, and priors before results from individual analyses are trusted.

Next, we provide a brief introduction to Bayesian analyses and Bayes factors, before we report the methods and results of our simulation studies.

\section{Introduction to Bayesian analyses, Bayes factors, and simulation-based calibration}\label{introduction-to-bayesian-analyses-bayes-factors-and-simulation-based-calibration}

Here, we provide a brief introduction to Bayesian analyses, Bayes factors, and to marginal SBC for Bayes factors. This introduction is taken from Schad et al. (2024). Other introductory treatments are available for Bayesian and Bayes factor analyses (Chow \& Hoijtink, 2017; Etz, Gronau, Dablander, Edelsbrunner, \& Baribault, 2018; Etz \& Vandekerckhove, 2018; Hoijtink \& Chow, 2017; Lee, 2011; Mulder \& Wagenmakers, 2016; Nicenboim, Schad, \& Vasishth, 2025; Nicenboim \& Vasishth, 2016; {van Doorn et al.}, 2021; Vandekerckhove et al., 2018; Vasishth, Nicenboim, Beckman, Li, \& Kong, 2018).

\subsection{A quick review of Bayesian methodology}\label{a-quick-review-of-bayesian-methodology}

Statistical analyses in the cognitive sciences often pursue two goals: to estimate parameters and their uncertainties, and to test hypotheses. Both of these goals can be achieved using Bayesian data analysis. Bayesian analyses focus on an ``observational'' model \(\mathcal{M}\), which specifies the probability density of the data \(y\) given the vector of model parameters \(\Theta\) and the model \(\mathcal{M}\), i.e., \(p(y \mid \Theta, \mathcal{M})\), or by dropping the model, \(p(y \mid \Theta)\). It is possible to use the observational model to simulate data, by selecting some model parameters \(\Theta\) and drawing random samples for the data \(\tilde{y}\). When the data is given (fixed, e.g., observed or simulated), then the observational model turns into a likelihood function: \(p(y \mid \Theta) = L_y(\Theta)\); this can be used to estimate model parameters or to compute the evidence for the model relative to other models. To estimate parameters, in Bayesian data analyses the likelihood is complemented by the prior, written \(p(\Theta)\), that defines the joint probability distribution of all the parameters of the model, which can encode domain expertise. Bayes' rule specifies how to combine the likelihood and the prior to compute the posterior distribution of the model parameters \(p(\Theta \mid y, \mathcal{M}_1)\):

\begin{equation}
p(\Theta \mid y, \mathcal{M}_1) = \frac{p(y \mid \Theta, \mathcal{M}_1) p(\Theta \mid \mathcal{M}_1)}{p(y \mid \mathcal{M}_1)} \label{eq:marginall}
\end{equation}

\noindent
Here, \(p(y \mid \mathcal{M}_1)\) is a normalizing constant termed the ``evidence'' or ``marginal likelihood''; it is the likelihood of the data \(y\) based on the model \(\mathcal{M}_1\) averaging over the parameters \(\Theta\). The marginal likelihood can only be interpreted relative to other models. It is derived as \(p(y \mid \mathcal{M}_1) = \int p(y \mid \Theta, \mathcal{M}_1) p(\Theta \mid \mathcal{M}_1) \, d \Theta\). There is a key role for priors in this computation. Priors play an important role in Bayesian inference
since expressing the prior is a means of quantifying the researcher's uncertainty about the model parameters. From a practical perspective, in Bayesian inference, priors can regularize inferences when the data do not strongly inform the likelihood functions. However, they will influence marginal likelihoods even when the data are strongly informative, and are thus even more crucial for Bayes factors than for posterior distributions (Aitkin, 1991; Gelman et al., 2013; Grünwald, 2000; Kass \& Raftery, 1995; Liu \& Aitkin, 2008; Myung \& Pitt, 1997; Ravenzwaaij \& Wagenmakers, 2022; Tendeiro \& Kiers, 2019; Vanpaemel, 2010).

For very simple models, posterior density functions can be computed analytically. However, for most interesting models this is not possible and we have to rely on approximation methods. One such approach is sampling methods such as Markov Chain Monte Carlo sampling (Gamerman \& Lopes, 2006), which is an important method behind popular software implementing Bayesian analysis.

\paragraph{Inference over hypotheses}

Bayes factors provide a way to compare any two model hypotheses (i.e., arbitrary hypotheses) against each other by comparing their marginal likelihoods (Betancourt, 2018; Kass \& Raftery, 1995; Ly, Verhagen, \& Wagenmakers, 2016; Schad et al., 2023). The Bayes factor tells us, given the data and the model priors, how much we need to update our relative belief between the two models.

To derive Bayes factors, we first compute the model posterior, i.e., the posterior probability for a model \(\mathcal{M}_i\) given the data:

\begin{equation}
p(\mathcal{M}_i \mid y) = \frac{p(y \mid \mathcal{M}_i) \times p(\mathcal{M}_i)}{p(y \mid \mathcal{M}_1) \times p(\mathcal{M}_1) + p(y \mid \mathcal{M}_2) \times p(\mathcal{M}_2)}= \frac{p(y \mid \mathcal{M}_i) \times p(\mathcal{M}_i)}{p(y)} .
\end{equation}

Here, \(p(\mathcal{M}_i)\) is the prior probability for each model \(i\). Based on the posterior model probability \(p(\mathcal{M}_i \mid y)\), we can compute the model odds for one model over another as:

\begin{equation}
\frac{p(\mathcal{M}_1 \mid y)}{p(\mathcal{M}_2 \mid y)} = \frac{[p(y \mid \mathcal{M}_1) \times p(\mathcal{M}_1)] / p(y)}{[p(y \mid \mathcal{M}_2) \times p(\mathcal{M}_2)] / p(y)} = \frac{p(y \mid \mathcal{M}_1)}{p(y \mid \mathcal{M}_2)} \times \frac{p(\mathcal{M}_1)}{p(\mathcal{M}_2)}\;. \label{eq:PostRatio}
\end{equation}

In words:

\begin{equation}
Posterior\;ratio = Bayes\;factor \times prior\;odds\;.
\end{equation}

The Bayes factor is thus a measure of relative evidence, the comparison of the predictive performance of one model (\(\mathcal{M}_1\)) against another one (\(\mathcal{M}_2\)). This comparison (\(BF_{12}\)) is a ratio of marginal likelihoods:

\begin{equation}
BF_{12} = \frac{p(y \mid \mathcal{M}_1)}{p(y \mid \mathcal{M}_2)}\;.
\end{equation}

\(BF_{12}\) indicates the evidence that the data provide for \(\mathcal{M}_1\) over \(\mathcal{M}_2\); in other words, which of the two models is more likely to have generated the data, or the relative evidence that we have for \(\mathcal{M}_1\) over \(\mathcal{M}_2\). Bayes factor values larger than one indicate that \(\mathcal{M}_1\) is more compatible with the data, values smaller than one indicate \(\mathcal{M}_2\) is more compatible with the data, and values close to one indicate that both models are equally compatible with the data.

In the present work, we will consider the case of nested model comparison. Here, a null model hypothesizes that a model parameter is zero or absent (a point hypothesis). By contrast, an alternative model hypothesizes that the model parameter is present with some prior distribution and has some value different from exactly zero (a ``general'' hypothesis).

As mentioned above, for most interesting problems and models in cognitive science, Bayes factors cannot be computed analytically; approximations are needed. One major approach is to estimate Bayes factors based on posterior MCMC draws via bridge sampling (Bennett, 1976; Gronau et al., 2017; Meng \& Wong, 1996), implemented in the R package \texttt{bridgesampling} (Gronau et al., 2020), which we use in the present work. We chose this package for the present study because it is widely used in cognitive science, and because it allows for straightforward application of marginal SBC. We did not use the Bayes factor package for analysis. Simulation-based calibration requires simulating from the parameter priors, which is not straightforward to implement for the Bayes factor package, due to its use of improper prior distributions.

\paragraph{Introduction to marginal simulation-based calibration (SBC)}

We have recently used marginal SBC for Bayes factors (Schad et al., 2023), which is a statistical technique designed to test whether a Bayes factor estimated in a given analysis is accurate, or whether it is biased and deviates from the true Bayes factor. Here, we first provide a short description of marginal SBC (derived from Schad et al., 2023).

We can formulate marginal SBC for Bayes factors (i.e., for model inference), where \(\mathcal{M}\) is a true model used to simulate artificial data \(y\), and \(\mathcal{M}'\) is a model inferred from the simulated data. Marginal SBC for model inference makes use of model priors p(\(\mathcal{M}\)), which are the prior probabilities for each model before observing any data. Marginal SBC can then be formulated as follows (Schad et al., 2023):

\begin{equation}
p(\mathcal{M}') = \sum_{\mathcal{M}} \int p(\mathcal{M}' \mid y) p(y \mid \mathcal{M}) p(\mathcal{M}) \, \mathrm{d} y \label{eq:SBC} \;.
\end{equation}

We can read this equation sequentially (from right to left):
first, we sample a model from the model prior, \(p(\mathcal{M})\). Next, we simulate data based on this model, \(p(y \mid \mathcal{M})\). This in fact involves two steps: simulating parameters from the parameter prior, \(p(\Theta \mid \mathcal{M})\), and then simulating artificial data from the parameters and model, \(p(y \mid \Theta, \mathcal{M})\), i.e., \(p(y \mid \mathcal{M}) = \int p(y \mid \Theta, \mathcal{M}) \times p(\Theta \mid \mathcal{M}) \, \mathrm{d} \Theta\). Next, we estimate the posterior model probabilities from the simulated data, \(p(\mathcal{M}' \mid y)\). This again involves several steps: we compute marginal likelihoods, \(p(y \mid \mathcal{M}')\), for both models, compute Bayes factors, and compute the posterior probability for each model given the data, \(p(\mathcal{M}' \mid y)\), by adding the model prior.
That is, we obtain a posterior model probability for each simulated data set. We can now compare, whether the posterior model probability, \(p(\mathcal{M}' \mid y)\), -- averaged across all simulated data sets -- is the same as the prior model probability, \(p(\mathcal{M})\).

The key idea is that if the computation of Bayes factors and posterior model probabilities is performed correctly (and of course the data simulation is implemented correctly), then the average posterior probability for a model should be the same as its prior probability. By contrast, if the average posterior probability for a model deviates from its prior probability, then this indicates that the Bayes factor estimate is biased, i.e., that the obtained Bayes factor does not correspond to the true Bayes factor.

For a given run of marginal SBC, to test whether the Bayes factor estimates are biased, we perform Bayesian t-tests on the posterior model probabilities, i.e., computing null hypothesis Bayes factors to test whether the posterior model probabilities differ from the prior model probabilities. Specifically, \textcolor{black}{for tests involving all simulations, we perform one-sample Bayesian t-tests, where for each simulation, we compare the posterior model probability (taking values between 0 and 1) against the true prior model probability (taking values of either 0 or 1). Moreover, for tests involving subsets of simulations (e.g., only simulations with a warning message),} we perform paired Bayesian t-tests, where for each simulation, we compare the posterior model probability (taking values between 0 and 1) against the prior sampled model (taking values of either 0 or 1). We test the two hypotheses that the difference between prior and posterior is zero (H0), or is not zero (H1) and that the standardized effect size of the difference follows a Cauchy distribution with default scale parameter of \(\sqrt{2}/2\). We moreover perform sensitivity analyses, where we vary the scale parameter to study its influence on the results.
We perform these t-tests using the R-package \texttt{BayesFactor} (function \texttt{ttestBF}) to test whether SBC indicates bias in the estimation of Bayes factors.

We note that the marginal SBC described here (i.e.~that data-averaged posterior equals the prior) is a different check than what is commonly referred to as SBC in the literature, (sometimes called \emph{conditional} SBC to make the difference explicit). See Modrák et al. (2023) for a comparison between the marginal and conditional formulation as well as review of other related approaches to Bayesian model checking. Conditional SBC is sensitive to different computation failures than marginal SBC. A related desirable property of Bayes factor is that the posterior model probabilities are calibrated, i.e.~that whenever \(p(\mathcal{M}_i \mid y) = q\), the true model should be \(\mathcal{M}_i\) in \(q\) of the cases. This calibration can be directly tested with a range of approaches, here we will use diagrams described in (Dimitriadis, Gneiting, \& Jordan, 2021). In fact, this form of calibration is equivalent to conditional SBC in a suitable reformulation of the problem. Proof of the equivalence and more discussion can be found in (Modrák et al., 2025).

\section{Methods}\label{methods}

We performed marginal SBC analyses for a range of experimental designs. Here, we report the methodological details of these simulation analyses.

\subsection{Design 1: 2x2 design with random effects for subjects}\label{design-1-2x2-design-with-random-effects-for-subjects}

We first performed marginal SBC for Bayes factors in a \(2 \times 2\) design inspired from the 2-step decision-making task (Daw, Gershman, Seymour, Dayan, \& Dolan, 2011). In this task, the probability to repeat an action is examined as a function of whether the last trial was rewarded and whether during the last trial a common or a rare transition occurred in the task. The task is designed to investigate the influence of model-free versus model-based reinforcement learning strategies on human learning and decision-making, where a model-free strategy is captured by the main effect of reward, whereas a model-based strategy is captured by the interaction reward \(\times\) transition probability. Here, we assume that the decisions to repeat an action are averaged for each of several blocks of trials, yielding an approximation to a normal distribution.

In the \(2 \times 2\) repeated measures design, each of the \(4\) conditions was repeated \(5\) times/blocks, leading to \(20\) (averaged) data points per subject. We simulated data from \(10\) subjects, yielding a total of \(20 \times 10 = 200\) data points.
We simulated data from the following model (H1), where \(n\) indicates the row of the data frame, and \(subj[n]\) yields the subject number in row \(n\).

\begin{align}
y_{n} &\sim Normal(\mu_{n}, \sigma_{Residual}) \\
\mu_{n} = &\beta_{(Intercept)} + u_{subj[n],1} + \\
        &(\beta_{meA} + u_{subj[n],2}) \times \text{meA}_{n} + \\
        &(\beta_{meB} + u_{subj[n],3}) \times  \text{meB}_{n} + \\
        &(\beta_{int} + u_{subj[n],4}) \times  \text{int}_{n}
\end{align}

The random effects \(u\) are multivariate normally distributed:

\begin{equation}\label{eq:jointpriordist1}
\begin{pmatrix}
  u_1 \\ 
  u_2 \\
  u_3 \\
  u_4 \\
\end{pmatrix}
\sim 
\mathcal{N} \left(
\begin{pmatrix}
  0 \\
  0 \\
  0 \\
  0 \\
\end{pmatrix},
\Sigma_{u}
\right),
\end{equation}

\noindent
where \(\Sigma_{u}\) is the variance-covariance matrix, with the standard deviation parameters \(\sigma_{Random\,slopes}\) and the correlation parameters \(\rho_{Random\,slopes}\).
The contrast coding for the two main effects and the interaction (\(\text{meA}_{n}\), \(\text{meB}_{n}\), and \(\text{int}_{n}\)) was set to sum-coding (\(-1\)/\(+1\)).
The model parameters used for simulating the SBC data were inspired by our own data on the task ({Schad et al.}, 2014). The prior distributions (from which simulating parameter values were sampled) were the following:

\begin{align}
\beta_{(Intercept)} &\sim Normal(0.7, 0.1) \\
\beta_{Fixed\,effects} &\sim Normal(0, 0.1) \\
\sigma_{Random\,slopes} &\sim Normal_+(0, 0.1) \\
\sigma_{Residual} &\sim Normal_+(0, 0.1) \\
\rho_{Random\,slopes} &\sim LKJ(2)
\end{align}

\noindent
where \(Normal_+\) denotes a normal distribution truncated at zero.

We performed three SBC simulations, one for main effect A, one for main effect B, and one for the interaction A \(\times\) B. Each simulation used the same H1 model (shown above). In each simulation, the null model (H0) was created by setting one fixed effects parameter to zero. When performing SBC for main effect A, \(\beta_{meA} = 0\); when performing SBC for main effect B, \(\beta_{meB} = 0\); when performing SBC for the interaction, \(\beta_{int} = 0\).
In each SBC simulation, we defined prior model probabilities as \(p(H0) = 0.5\) and \(p(H1) = 0.5\).
In each SBC simulation, we fitted the same models to the data that were used to generate the data (i.e., H1 and the corresponding H0). In the model fitting, we used \(10,000\) iterations of the MCMC sampler, with \(2,000\) iterations of warm-up. For the bridge sampling, we used the ``warp3'' method. We performed \(200\) runs of SBC.

\subsection{Design 2: One 2-level factor with crossed random effects for subjects and items}\label{design-2-one-2-level-factor-with-crossed-random-effects-for-subjects-and-items}

We simulated data from a Latin square design with one two-level fixed factor and crossed random effects for \(42\) subjects and \(16\) items, leading to a total of \(42 \times 16 = 672\) data points. We set the prior probability of the alternative hypothesis (H1) to a smaller value of \(0.2\) (instead of the previous \(0.5\)) to be able to detect potential liberal biases well without ceiling effects.
A Bayesian GLMM with a lognormal likelihood was used in the SBC analyses using the package brms. We included correlated random slopes and intercepts for subjects and for items.

\begin{align}
y_{n} &\sim LogNormal(\mu_{n}, \sigma_{Residual}) \\
\mu_{n} = &\beta_{(Intercept)} + u_{subj[n],1} + w_{item[n],1} + \\
         &(\beta_{X}           + u_{subj[n],2} + w_{item[n],2}) \times X_{n}
\end{align}

The random effects \(u\) and \(w\) are each multivariate normally distributed:

\begin{equation}
\begin{pmatrix}
  u_1 \\ 
  u_2 \\
\end{pmatrix}
\sim 
\mathcal{N} \left(
\begin{pmatrix}
  0 \\
  0 \\
\end{pmatrix},
\Sigma_{u}
\right),
\end{equation}

\begin{equation}
\begin{pmatrix}
  w_1 \\ 
  w_2 \\
\end{pmatrix}
\sim 
\mathcal{N} \left(
\begin{pmatrix}
  0 \\
  0 \\
\end{pmatrix},
\Sigma_{w}
\right),
\end{equation}

\noindent
where \(\Sigma_{u}\) and \(\Sigma_{w}\) are the variance-covariance matrices, with the standard deviation parameters \(\sigma_{Random\,slopes}\) and the correlation parameters \(\rho_{Random\,slopes}\).
The contrast coding for the fixed effect (\(X_{n}\)) was set to sum-coding (\(-1\)/\(+1\)).
The prior distributions (from which simulating parameter values were sampled) were the following:

\begin{align}
\beta_{(Intercept)} &\sim Normal(6, 0.6) \\
\beta_{X} &\sim Normal(0, 0.15) \\
\sigma_{Random\,slopes} &\sim Normal_+(0, 0.1) \\
\sigma_{Residual} &\sim Normal_+(0, 0.5) \\
\rho_{Random\,slopes} &\sim LKJ(2)
\end{align}

We performed one SBC simulation comparing this full model (H1) to a null model (H0), where the fixed effect was set to zero: \(\beta_{X} = 0\).
In the SBC, we fitted the same models to the data that were used to generate the data (i.e., H1 and H0). In the model fitting, we used \(10,000\) iterations, with \(2,000\) iterations of warm-up. For bridge sampling, we again used the method ``warp3'' with the aim to reduce the number of warning messages.

\subsection{Design 3: 2x2 repeated measures design with crossed random effects for subjects and items}\label{design-3-2x2-repeated-measures-design-with-crossed-random-effects-for-subjects-and-items}

Next, we performed SBC for Bayes factors in a \(2 \times 2\) Latin square design, with crossed random effects for subjects and items. We simulated data from \(16\) subjects and from \(8\) trials, leading to a total of \(16 \times 8 = 128\) data points.
We simulated data from the following model (H1), where \(n\) indicates the row of the data frame, \(subj[n]\) yields the subject number in row \(n\), and \(item[n]\) yields the item number in row \(n\).

\begin{align}
y_{n} &\sim Normal(\mu_{n}, \sigma_{Residual}) \\
\mu_{n} = &\beta_{(Intercept)} + u_{subj[n],1} + w_{item[n],1} + \\
        &(\beta_{meA} + u_{subj[n],2} + w_{item[n],2}) \times \text{meA}_{n} + \\
        &(\beta_{meB} + u_{subj[n],3} + w_{item[n],3}) \times  \text{meB}_{n} + \\
        &(\beta_{int} + u_{subj[n],4} + w_{item[n],4}) \times  \text{int}_{n}
\end{align}

The random effects \(u\) and \(w\) are each multivariate normally distributed:

\begin{equation}
\begin{pmatrix}
  u_1 \\ 
  u_2 \\
  u_3 \\
  u_4 \\
\end{pmatrix}
\sim 
\mathcal{N} \left(
\begin{pmatrix}
  0 \\
  0 \\
  0 \\
  0 \\
\end{pmatrix},
\Sigma_{u}
\right), 
\end{equation}

\begin{equation}
\begin{pmatrix}
  w_1 \\ 
  w_2 \\
  w_3 \\
  w_4 \\
\end{pmatrix}
\sim 
\mathcal{N} \left(
\begin{pmatrix}
  0 \\
  0 \\
  0 \\
  0 \\
\end{pmatrix},
\Sigma_{w}
\right),
\end{equation}

\noindent
where \(\Sigma_{u}\) and \(\Sigma_{w}\) are the variance-covariance matrices, each with the standard deviation parameters \(\sigma_{Random\,slopes}\) and the correlation parameters \(\rho_{Random\,slopes}\).
The contrast coding for the two main effects and the interaction (\(\text{meA}_{n}\), \(\text{meB}_{n}\), and \(\text{int}_{n}\)) was set to sum-coding (\(-1\)/\(+1\)).
The prior distributions (from which simulating parameter values were sampled) were the following:

\begin{align}
\beta_{(Intercept)} &\sim Normal(0.7, 0.1) \\
\beta_{Fixed\,effects} &\sim Normal(0, 0.1) \\
\sigma_{Random\,slopes} &\sim Normal_+(0, 0.1) \\
\sigma_{Residual} &\sim Normal_+(0, 0.1) \\
\rho_{Random\,slopes} &\sim LKJ(2)
\end{align}

\noindent
where \(Normal_+\) denotes a normal distribution truncated at zero.

We performed three SBC simulations, one for main effect A, one for main effect B, and one for the interaction A \(\times\) B. Each simulation used the same H1 model (shown above). In each simulation, the null model (H0) was created by setting one fixed effects parameter to zero. When performing SBC for main effect A, \(\beta_{meA} = 0\); when performing SBC for main effect B, \(\beta_{meB} = 0\); when performing SBC for the interaction, \(\beta_{int} = 0\).
In each SBC simulation, we defined prior model probabilities as \(p(H0) = 0.8\) and \(p(H1) = 0.2\).
In each SBC simulation, we fitted the same models to the data that were used to generate the data (i.e., H1 and the corresponding H0). In the model fitting, we used \(10,000\) iterations, with \(2,000\) iterations of warm-up. For bridge sampling, we used the ``warp3'' method. We performed \(200\) runs of SBC.
In additional simulations, we repeated the SBC analyses with \(40,000\) iterations, again with \(2,000\) warm-up iterations. Moreover, we performed SBC analysis with \(1,000\) SBC simulations (\(10,000\) MCMC samples).
\textcolor{black}{Last, we performed a simulation with $p(H0) = 0.2$ and $p(H1)=0.8$.}

\subsection{Selective reporting}\label{selective-reporting}

As is widely noted, Bayesian posteriors do not change under data-dependent selective reporting, i.e.~for any binary random variable \(A\), such that \(A\) is conditionally independent of \(\theta\) given \(y\), we have \(p(\theta | y) = p(\theta | y, A = 1)\). This implies that the expected posterior does not change when we filter datasets and therefore all consistency guarantees, including SBC, are valid also for just a subset of the simulations as long as the exclusion criterion is solely a function of the simulated data. Notably this includes any diagnostics from the Bayes factor computation as those only depend on data and it is therefore valid to perform SBC separately for simulations that satisfy a diagnostic check and simulations that fail a diagnostic check.

\section{Results}\label{results}

\subsection{Design 1: 2x2 design with random effects for subjects}\label{design-1-2x2-design-with-random-effects-for-subjects-1}

\begin{figure}

{\centering \includegraphics[width=\textwidth]{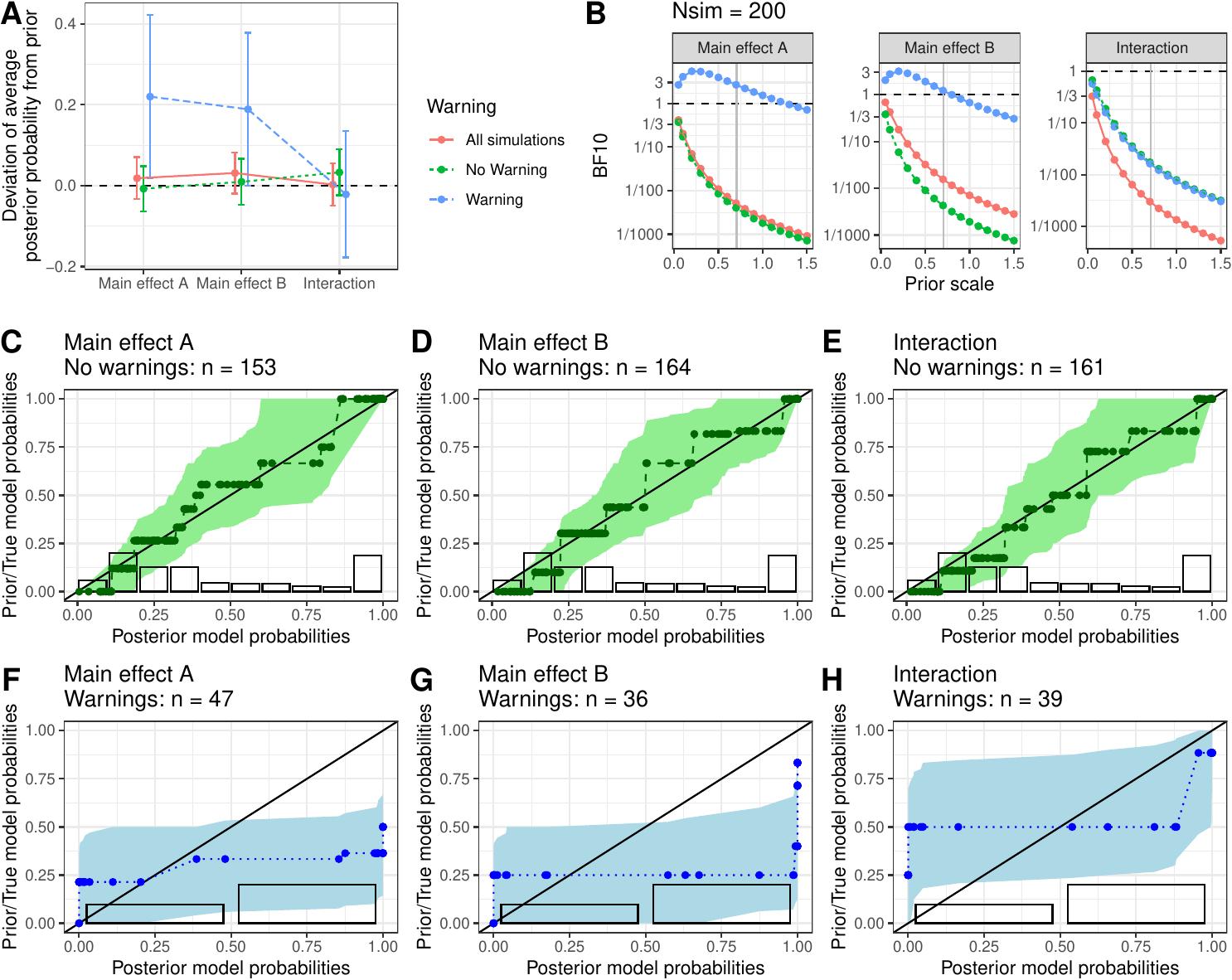} 

}

\caption{Results from simulation-based calibration (SBC) for a 2 x 2 repeated measures design. A) The average deviation of the posterior model probability for H1 from the prior (with 95\% confidence intervals) is shown for the main effects A, B, and their interaction. Data is shown for all simulations and separately for those without / with a warning message from the bridge sampling algorithm. The horizontal dashed line indicates no deviation; deviations from this line indicate bias. B) Bayesian t-tests of whether the posterior probability for H1 deviates from the prior probability, with sensitivity analyses for the prior scale. The vertical grey line indicates a default Bayes factor. C-H) Reliability diagrams: the prior model probability is plotted as a function of the estimated posterior model probabilities. Error bands show 95\% consistency bands. Results show that when no warning message was issued (C-E) prior model probabilities do not deviate from posterior model probabilities, providing evidence for accurate estimation, but that when a warning message was issued (F-H) prior and posterior model probabilities do not match, providing evidence for biases.}\label{fig:2step-plot}
\end{figure}

First, we studied a repeated measures \(2 \times 2\) design, which was inspired by the 2-step decision-making task (Daw et al., 2011; {Schad et al.}, 2014), by running marginal SBC. We investigated model convergence. Across all model fits (\(200\) simulations \(\times 3\) effects \(\times 2\) models (H0/H1) \(= 1200\) model fits), we observed 0 divergent transitions. R-hat exceeded the threshold of 1.01 for at least one population-level parameter in 0
cases, suggesting good model convergence.

However, during the bridge sampling, there was the warning message ``logml could not be estimated within maxiter, rerunning with adjusted starting value. Estimate might be more variable than usual'' in 122 simulations (20.33\%), suggesting problems with the bridge sampling. Due to these problems, we looked at the results separately for simulations with versus without warning messages from the bridge sampler.

The results for marginal SBC (see Fig.~\ref{fig:2step-plot}A-B) showed for simulations without warning messages that the posterior model probabilities were equal to the prior model probability of \(0.5\) for both main effects and for the interaction. Indeed, the null hypothesis that posterior model probabilities were equal to \(0.5\) was supported by Bayesian t-tests (see Fig.~\ref{fig:2step-plot}B) showing default Bayes factors of \(BF01 > 10\) for the null for the two main effects and the interaction. A prior sensitivity analysis showed that these results were stable across different prior scales, being present already for very small prior scales, and becoming stronger for larger prior scales.
These results suggest no evidence for overall bias in the Bayes factor estimates when no warning message is shown.
For simulations where there was a warning message from the bridge sampler, the results were less clear, with sensitivity analyses showing either evidence for no bias (interaction), or unclear evidence (main effects).

We moreover used reliability diagrams (Dimitriadis et al., 2021) to investigate whether the distribution of estimated posterior model probabilities is calibrated. If the estimates are well calibrated and accurate, posterior model probabilities should accurately predict the true model, which would be visible in Figure~\ref{fig:2step-plot}C-H in estimates aligned to the diagonal. Indeed, it is visible in Figure~\ref{fig:2step-plot}C-E that when no warning message was issued by the bridge sampler prior and posterior model probabilities did match, suggesting accurate estimation. However, in cases where the bridge sampling algorithm issued a warning message, the estimates deviated from the diagonal, suggesting bias in the resulting Bayes factor estimate (see Fig.~\ref{fig:2step-plot}F-H).

Importantly, additional analyses showed that we observed warnings from the bridgesampling package only for data sets that required small step sizes and large treedepths to explore with Stan's HMC sampler and which in turn required a lot of computation time to fit. This implies that the posterior geometry was highly irregular, which would explain problems faced by bridge sampling. In particular, the problematic data sets tended to result in small fitted residual standard deviations and/or some random effect standard deviations close to zero. When faced with such posteriors, a user may consider reparametrizing the model to avoid the computational difficulties and we would expect a successful reparametrization to also resolve the bridgesampling warnings. E.g. for the cases with low residual standard deviations, the data are actually highly informative of the random effects which results in weak non-identifiability between the overall intercept/effect and random intercepts/effects (see Ogle \& Barber, 2020, for more discussion). The models worked well when dropping the random effect for the interaction. Also, in some (but not all) of the cases, enforcing a sum-to-zero constraint on the random intercept and each random effect results in improved sampling and no warnings from bridgesampling.

\subsection{Design 2: One 2-level factor with crossed random effects for subjects and items}\label{design-2-one-2-level-factor-with-crossed-random-effects-for-subjects-and-items-1}

\begin{figure}

{\centering \includegraphics[width=\textwidth]{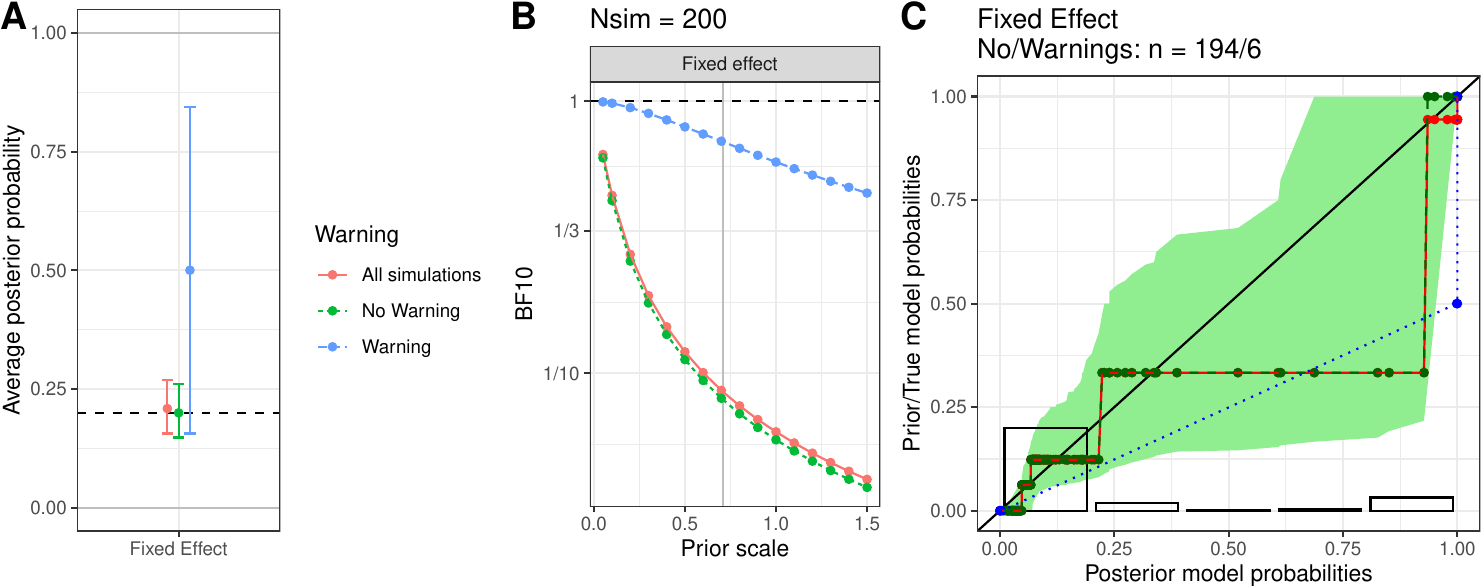} 

}

\caption{Latin square design with one 2-level fixed factor and random effects for subjects and items. A) The average posterior model probability for H1 (with 95\% confidence intervals) is shown for the fixed effect. Data is shown for all simulations and separately for those without / with a warning message from the bridge sampling algorithm. The horizontal dashed line is the prior probability for the H1; deviations from this line indicate bias. B) Bayesian t-tests of whether the posterior probability for H1 deviates from the prior probability, with sensitivity analyses for the prior scale. The vertical grey line indicates a default Bayes factor, providing evidence for no bias. C) Reliability diagram: the prior model probability is plotted as a function of the estimated posterior model probabilities. Error bands show 95\% consistency bands (only for simulations without warning). Results show that prior model probabilities do not deviate from posterior model probabilities, providing evidence for accurate estimation.}\label{fig:ItemSO-plot}
\end{figure}

Next, we studied a Latin-square design with a 2-level fixed factor and crossed random effects for subjects and items, by running SBC. For the bridge sampling, we again used the method ``warp3'' with the aim to keep the number of warning messages of the bridge sampling algorithm low. We first investigated model convergence. Across all model fits (\(200\) simulations \(\times 2\) models (H0/H1) \(= 400\) fits), we observed 0 divergent transitions. R-hat exceeded the threshold of 1.01 for at least one population-level parameter in 0 cases, suggesting good model convergence.
During the bridge sampling, there was the warning message ``logml could not be estimated within maxiter, rerunning with adjusted starting value. Estimate might be more variable than usual'' in 6 simulations (3.00\%), suggesting only few problems with the bridge sampling. Therefore, we performed statistical analysis only for simulations without a warning message from the bridge sampler.

The results (see Fig.~\ref{fig:ItemSO-plot}) showed that the average posterior model probabilities for the H1 did not differ from the prior (marginal SBC). Indeed, default Bayes factor analyses with Bayes factors of \(BF01 > 10\) provided support for the null hypothesis that the average posterior probability for H1 did not differ from the prior of \(0.2\), and sensitivity analyses showed that this result was stable across a larger range of prior scales. The results from the reliability diagram (see Fig.~\ref{fig:ItemSO-plot}C) also did not provide evidence for bias, but rather suggested accurate Bayes factor estimation when no warning message was shown.

\subsection{Design 3: 2x2 repeated measures design with crossed random effects for subjects and items}\label{design-3-2x2-repeated-measures-design-with-crossed-random-effects-for-subjects-and-items-1}

\begin{figure}

{\centering \includegraphics[width=\textwidth]{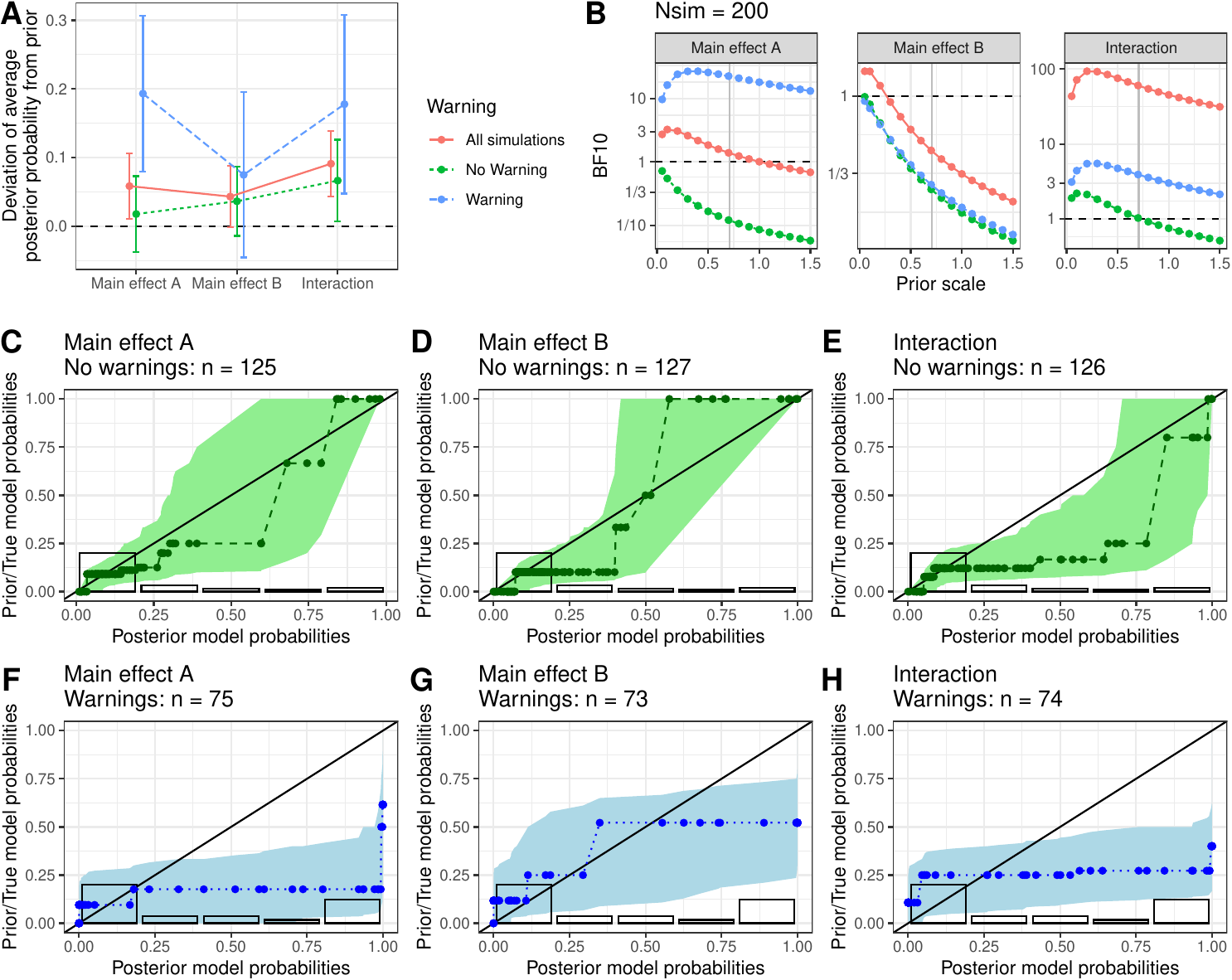} 

}

\caption{A 2 x 2 repeated measures design with crossed random effects for subjects and items. A) The average deviation of the posterior model probability for H1 from the prior (95\% CI) is shown for the main effects A, B, and their interaction. Data is shown for all simulations and separately for those without / with a warning message from the bridge sampling algorithm. The horizontal dashed line indicates not deviation; deviations from this line indicate bias. B) Bayesian t-tests of whether the posterior probability for H1 deviates from the prior, with sensitivity analyses for the prior scale. The vertical grey line indicates a default Bayes factor. C-H) Reliability diagrams: the prior model probability is plotted as a function of the estimated posterior model probabilities for simulations without (C-E) and with (F-H) a warning message. Error bands show 95\% consistency bands. Results show that when no warning is issued, prior model probabilities largely do not deviate from posterior model probabilities. Results for the interaction are unclear. However, biases are visible in simulations with a warning message.}\label{fig:2stepItem-plot}
\end{figure}

We studied a repeated measures \(2 \times 2\) Latin square design with crossed random effects for subjects and items. We again used the ``warp3'' method for bridge sampling, and used \(10,000\) iterations of MCMC sampling. We first investigated model convergence. Across all model fits (\(200\) simulations \(\times 3\) effects \(\times 2\) models (H0/H1) \(= 1200\) fits), we observed 0 divergent transitions. R-hat exceeded the threshold of 1.01 for at least one population-level parameter in 4 cases, suggesting overall good model convergence.
However, during the bridge sampling, there was the warning message ``logml could not be estimated within maxiter, rerunning with adjusted starting value. Estimate might be more variable than usual'' in 222 simulations (37.00\%), suggesting problems with the bridge sampling.
\textcolor{black}{In} 214 \textcolor{black}{ cases, the warning message occurred for the H0 model. For the H1 model, there were warning messages in} 209 \textcolor{black}{ cases. Out of the total $3 \times 200 = 600$ simulations,} 487 \textcolor{black}{simulated the H0 and} 113 \textcolor{black}{simulated the H1. From these simulations, of the H0 simulations,} 170 (35\%) \textcolor{black}{exhibited a warning message, and of the H1 simulations,} 52 (46\%) \textcolor{black}{exhibited a warning message, suggesting slightly increased percentage.
We again looked at the results separately for simulations with versus without warning messages from the bridge sampler.}

The results (see Fig.~\ref{fig:2stepItem-plot}) showed that -- for simulations without warning messages from the bridge sampling -- the posterior model probabilities did not differ from the prior model probability for the main effects. Indeed, the null hypothesis that the posterior model probability was equal to the prior was supported by Bayesian t-tests (see Fig.~\ref{fig:2stepItem-plot}B) showing Bayes factors of \(BF01 > 3\) for the null hypothesis for the two main effects (sensitivity analyses showed stronger effects for main effect A than B), and reliability diagrams did not show clear evidence for bias (Fig.~\ref{fig:2stepItem-plot}C-D). For the interaction, the results were not quite clear: the results from the Bayesian t-test strongly depended on the prior scale; moreover, a slight deviation seemed present in the reliability diagrams (see Fig.~\ref{fig:2stepItem-plot}E) and more SBC simulations may be needed to judge whether this deviation is reliable.

For simulations with a warning message, the results from the marginal SBC were mixed, with evidence for bias (\(BF10 > 3\)) in the case of main effect A and the interaction, but slight evidence for absence of bias in the case of main effect B (\(BF01 > 3\)). However, reliability diagrams (see Fig.~\ref{fig:2stepItem-plot}F-H) indicated rather clear evidence for miscalibration between posterior and prior probabilities, suggesting Bayes factor estimates were biased.

\begin{figure}

{\centering \includegraphics[width=\textwidth]{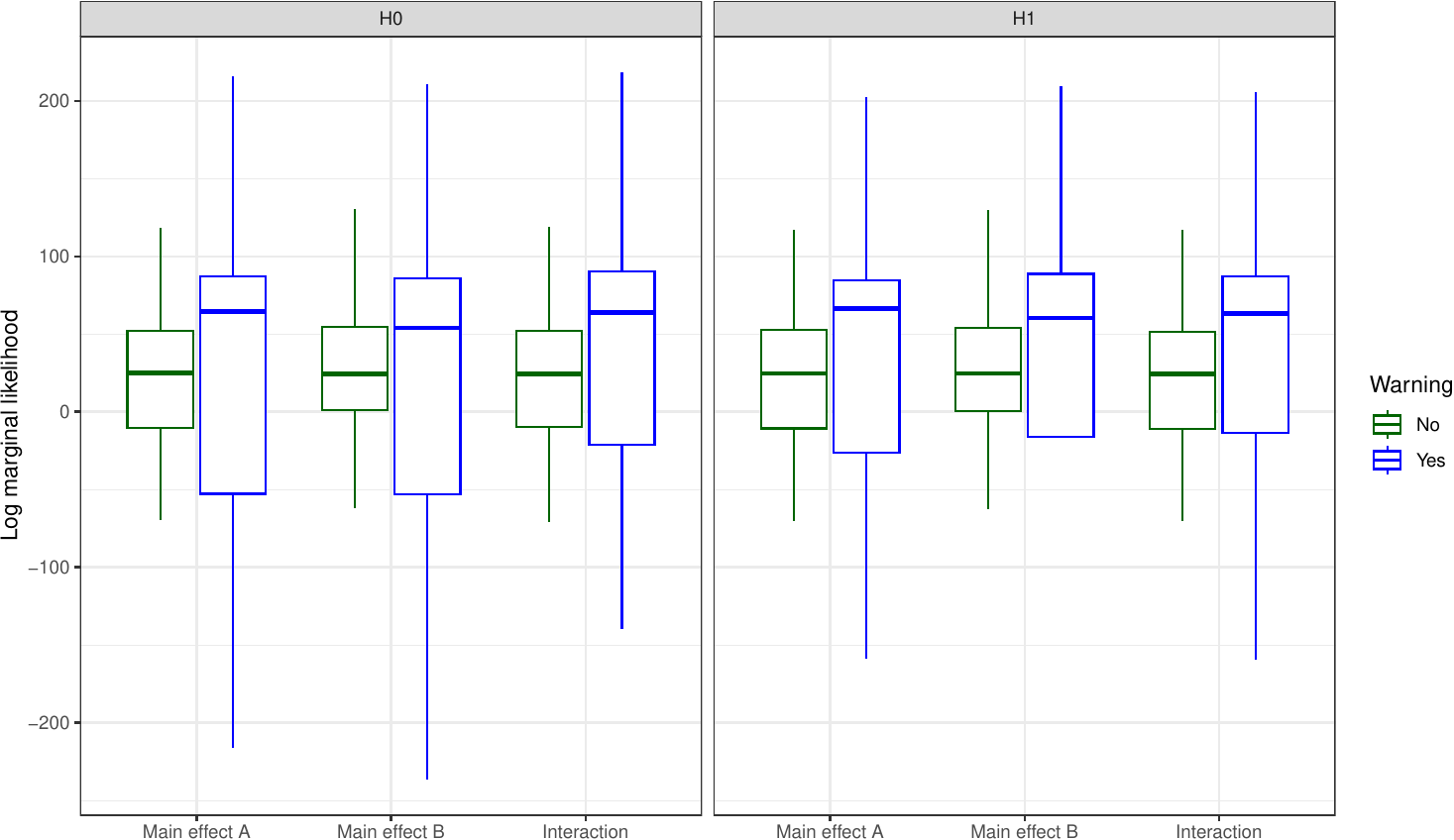} 

}

\caption{A 2 x 2 repeated measures design with crossed random effects for subjects and items: Box plots of the log marginal likelihoods for H0 and H1 models, for different effects, and for runs with versus without a warning message. Outliers removed in boxplots.}\label{fig:2stepItem-ML-plot}
\end{figure}

\textcolor{black}{We moreover checked the distributions of the log marginal likelihoods for the H1 and the H0 models across simulations. We found (see Fig.}~\ref{fig:2stepItem-ML-plot}) \textcolor{black}{that the variances were much enhanced when a warning message was issued by the bridge sampler.}

These results suggest evidence for absence of overall bias in the Bayes factor estimates for the main effects when no warning message is issued. The results for the interaction were unclear. However, in the cases where a warning message was issued by the bridge sampler, we saw evidence for bias in the Bayes factor estimates, with posterior model probabilities (for main effect A and for the interaction) being too large and larger than the prior probability (at least for large posterior model probabilities).

\subsubsection{Increasing the number of MCMC iterations to 40,000}\label{increasing-the-number-of-mcmc-iterations-to-40000}

The previous simulations suffered from a relatively large number of warning messages of the bridge sampler. With the aim to reduce this large number, we repeated the SBC simulations for the interaction term with \(40,000\) iterations of the MCMC sampler instead of the previous \(10,000\) iterations.

\begin{figure}

{\centering \includegraphics[width=\textwidth]{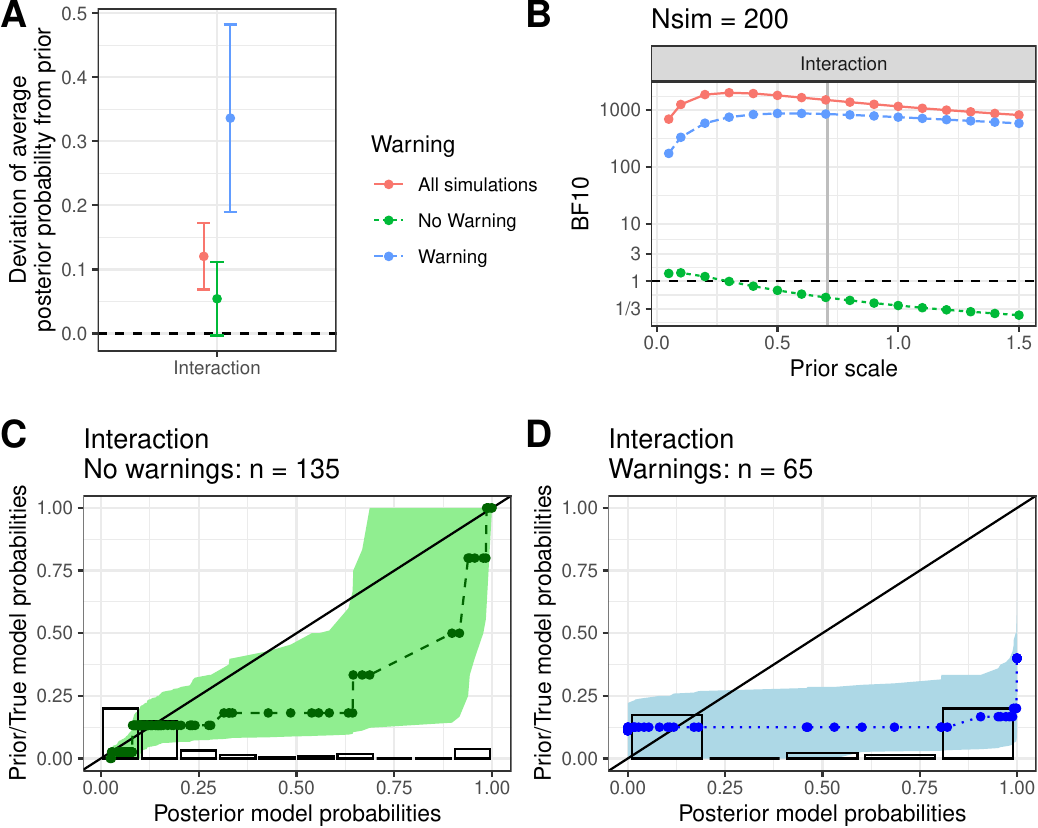} 

}

\caption{Repetition of the previous analyses of the interaction term  (in design 3), but with 40,000 iterations of MCMC sampling instead of 10,000. A) The average deviation of posterior model probabilities for H1 from the priors (95\% CI). All simulations and simulations without / with a warning message. Deviations from the horizontal dashed line indicate bias. B) Bayesian t-tests of whether the posterior probability for H1 deviates from the prior, with sensitivity analyses for the prior scale. The vertical grey line indicates a default Bayes factor. C+D) Reliability diagrams: the prior model probability is plotted as a function of the estimated posterior model probabilities for simulations without (C) and with (D) a warning message. Error bands show 95\% consistency bands. Results show that when no warning is issued, the results are unclear. When a warning message is issued, clear bias is visible.}\label{fig:2stepItem40000-plot}
\end{figure}

We repeated the previous simulations of design 3, but now only for the interaction term. We first investigated model convergence. Across all model fits (\(200\) simulations \(\times 2\) models (H0/H1) \(= 400\) fits), we observed 1 divergent transitions. R-hat exceeded the threshold of 1.01 for at least one population-level parameter in 0 cases, suggesting overall good model convergence.

However, during the bridge sampling, there was the warning message ``logml could not be estimated within maxiter, rerunning with adjusted starting value. Estimate might be more variable than usual'' in 65 simulations (32.50\%), suggesting warning messages were somewhat, but not strongly reduced. We again looked at the results separately for simulations with versus without warning messages from the bridge sampler.

The results (see Fig.~\ref{fig:2stepItem40000-plot}) showed that -- for simulations without warning messages from the bridge sampling -- it was not clear whether the posterior model probabilities differed from the prior model samples. Indeed, the null hypothesis that the posterior model probability was equal to the prior was neither clearly supported and nor clearly refuted by the Bayesian t-tests and the sensitivity analysis (see Fig.~\ref{fig:2stepItem40000-plot}B), but showed annecdotal evidence for no bias. However, in the cases where a warning message was issued by the bridge sampler, we saw evidence for bias in the Bayes factor estimates, with average posterior model probabilities being too large (and larger than the prior probability; \(BF10 > 3\)).

Moreover, reliability diagrams (see Fig.~\ref{fig:2stepItem40000-plot}C+D) again showed that for simulations without warning message it was unclear whether posterior model probabilities were the same as the prior probabilities. Specifically, for posterior model probabilities of around 50\%, the prior model probability tended to be slightly low, but the 95\% consistency bands did not provide clear evidence for bias. However, this was based on only very few simulations (see histogram in Fig.~\ref{fig:2stepItem40000-plot}C), and more SBC simulations may be needed to draw firm conclusions.
In simulations where a warning message was issued by the bridge sampler, the reliability diagrams suggested that posterior and prior model probabilities again did not match, indicating bias. Specifically, for high posterior probabilities (of H1) the prior probabilities were too low, again suggesting falsely high posterior model probabilities.

The results therefore suggest that increasing the number of MCMC iterations to \(40,000\) slightly reduced the proportion of warning message (from \(37\)\% to \(32.5\)\%), but this change is fully within the expected sampling variability for the size of the simulation study.

\subsubsection{Running 1000 SBC simulations}\label{running-1000-sbc-simulations}

The previous results are limited in that we only performed \(200\) simulations of marginal SBC. Here, for better stability and power, we performed \(1000\) SBC simulations of the interaction term in the \(2 \times 2\) repeated measures design using brms/bridgesampling.

\begin{figure}

{\centering \includegraphics[width=\textwidth]{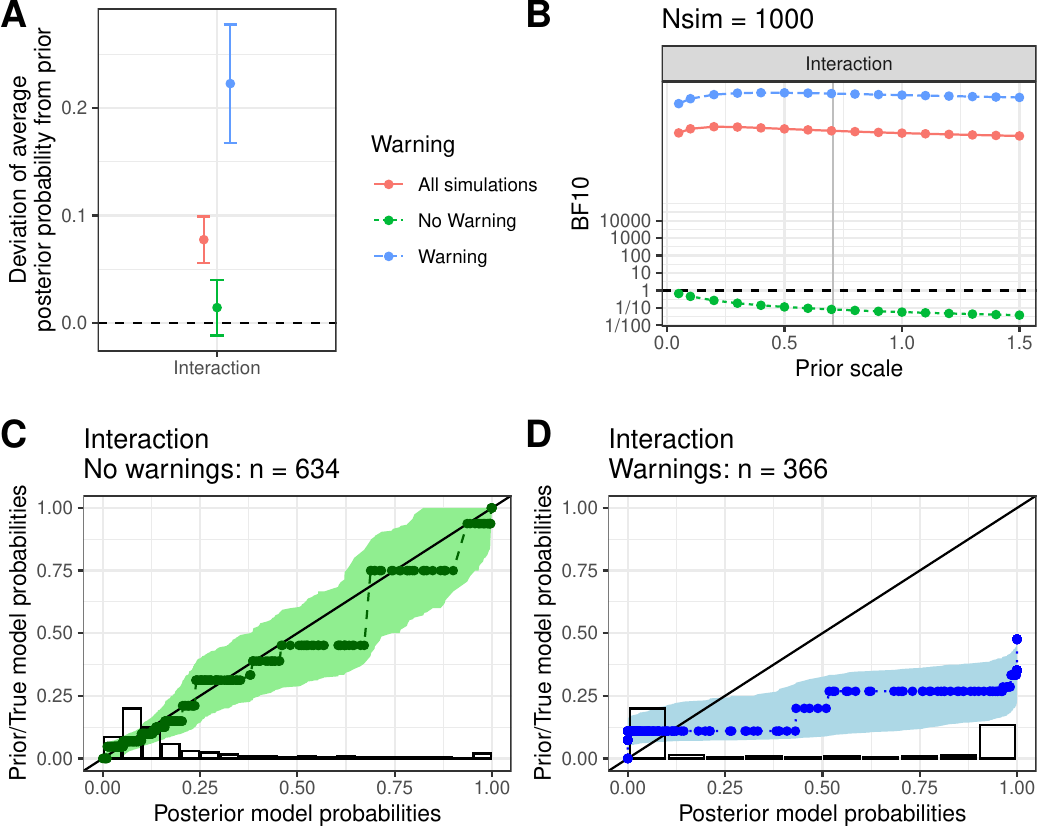} 

}

\caption{Repetition of the previous analyses of the interaction term  (in design 3), but with 1000 SBC simulations and 10,000 iterations of MCMC sampling. A) The average deviation of posterior model probabilities for H1 from the prior (95\% CI). All simulations and simulations without / with a warning message. The horizontal dashed line indicates absence of devation; deviations from this line indicate bias. B) Bayesian t-tests of whether the posterior probability for H1 deviates from the prior, with sensitivity analyses for the prior scale. The vertical grey line indicates a default Bayes factor. C+D) Reliability diagrams: the prior model probability is plotted as a function of the estimated posterior model probabilities for simulations without (C) and with (D) a warning message. Error bands show 95\% consistency bands. Results show that when no warning is issued prior model probabilities do not deviate from posterior model probabilities, suggesting accurate estimation, but that in simulations with a warning message they clearly deviate, providing evidence for bias in this case.}\label{fig:2stepItem1000it-plot}
\end{figure}

Analysis of model convergence showed that across all model fits (\(1000\) simulations \(\times 2\) models (H0/H1) \(= 2000\) fits), we observed 0 divergent transitions. R-hat exceeded the threshold of 1.01 for at least one population-level parameter in 2 cases, suggesting overall good model convergence.
However, during the bridge sampling, there was the warning message ``logml could not be estimated within maxiter, rerunning with adjusted starting value. Estimate might be more variable than usual'' in 366 simulations (36.60\%), again suggesting problems with the bridge sampling. We again looked at the results separately for simulations with versus without warning messages from the bridge sampler.

The results (see Fig.~\ref{fig:2stepItem1000it-plot}) showed that -- for simulations without warning messages from the bridge sampling -- the posterior model probabilities did not differ from the prior model samples. Indeed, the null hypothesis that the posterior model probability was equal to the prior was supported by Bayesian t-tests (see Fig.~\ref{fig:2stepItem1000it-plot}B) showing default Bayes factors of \(BF01 > 10\) for the null hypothesis, which was evident especially for larger prior scales. These results suggest evidence for absence of overall bias in the Bayes factor estimates when no warning message is issued. However, in the cases where a warning message was issued by the bridge sampler, we saw clear evidence for bias in the Bayes factor estimates, with average posterior model probabilities being too large (and larger than the prior probability; \(BF10 > 10000\)).

Moreover, reliability diagrams (see Figure~\ref{fig:2stepItem1000it-plot}C+D) showed that now - with sufficient SBC simulations - for the simulations where no warning message was shown, prior and posterior model probabilities did not deviate, suggesting no bias in these cases.
Moreover, in simulations where a warning message was issued by the bridge sampler, the reliability diagrams clearly showed that posterior and prior model probabilities did not match.

The results with larger numbers of SBC simulations therefore suggest that Bayes factor estimates are indeed accurate in simulations without a warning message, but are biased in simulations with a warning message.

\subsubsection{Results as a function of the number of samples}\label{results-as-a-function-of-the-number-of-samples}

In the previous analyses, we used different numbers of samples of SBC. As noted above, the number of SBC samples could impact the results since a larger number of samples may be needed to obtain enough power to detect divergences between the posterior and prior model probabilities, and to yield stable and reliable results. To investigate this issue, we next studied the results as a function of the number of SBC samples, i.e., performing a re-analysis of the previous marginal SBC simulations.

\begin{figure}

{\centering \includegraphics{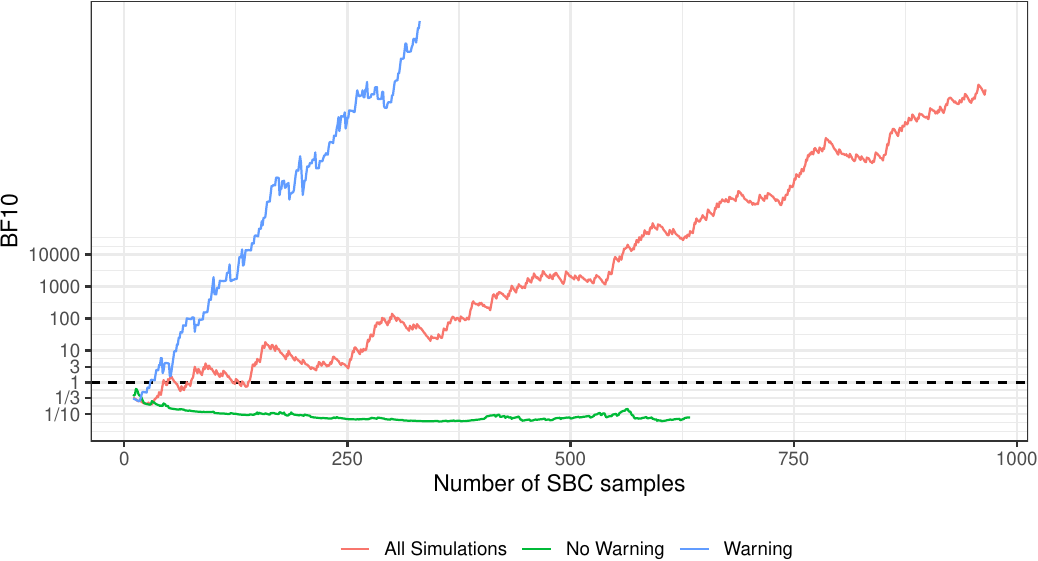} 

}

\caption{Analysis of how the number of SBC samples impacts marginal SBC. The result of a Bayesian default t-test of whether the average posterior model probability diverges from zero is plotted as a function of the number of SBC samples used in the analysis. Results are shown for all simulations (red) and for simulations without (green) and with (blue) a warning message. Evidence for bias in simulations with warning messages is visible very quickly, with only few SBC simulation runs, whereas evidence for accuracy without warning messages accumulates more slowly.}\label{fig:2stepItem1000it-nsamp-plot}
\end{figure}

The results showed (see Fig.~\ref{fig:2stepItem1000it-nsamp-plot}) that evidence for bias in simulations with warning messages was quickly evident with already less than 100 simulation runs, and became more and more clear with increasing runs. Evidence for accurate Bayes factor estimates in simulations without warning messages accumulated less quickly and remained at less clear levels with a Bayes factor of around \(BF01 = 10\).

\subsubsection{Using a prior model probability for H0 of 0.2 instead of 0.8}\label{using-a-prior-model-probability-for-h0-of-0.2-instead-of-0.8}

\textcolor{black}{The previous simulations were based on a situation, where the prior model probability for the H0 was 0.8 and the probability for H1 was set to 0.2. In this situation, we found that in simulations with a warning message, the average posterior model probability for H1 was too \textit{large}, i.e., larger than the prior model probability. As our last analysis, we aimed to test whether this originated from a general upward-bias of Bayes factors flagged with a warning message, or whether it originated from a more symmetric bias, e.g., a situation, where the average posterior was selected with a larger degree of noise (cf. our prior analysis showing the variance in the log marginal likelihoods was enhanced in simulations with a warning message). To this end, we performed the same simulations, but setting the prior probability for H0 to 0.2 instead of 0.8. We used $10,000$ iterations of the MCMC sampler.}

\begin{figure}

{\centering \includegraphics[width=\textwidth]{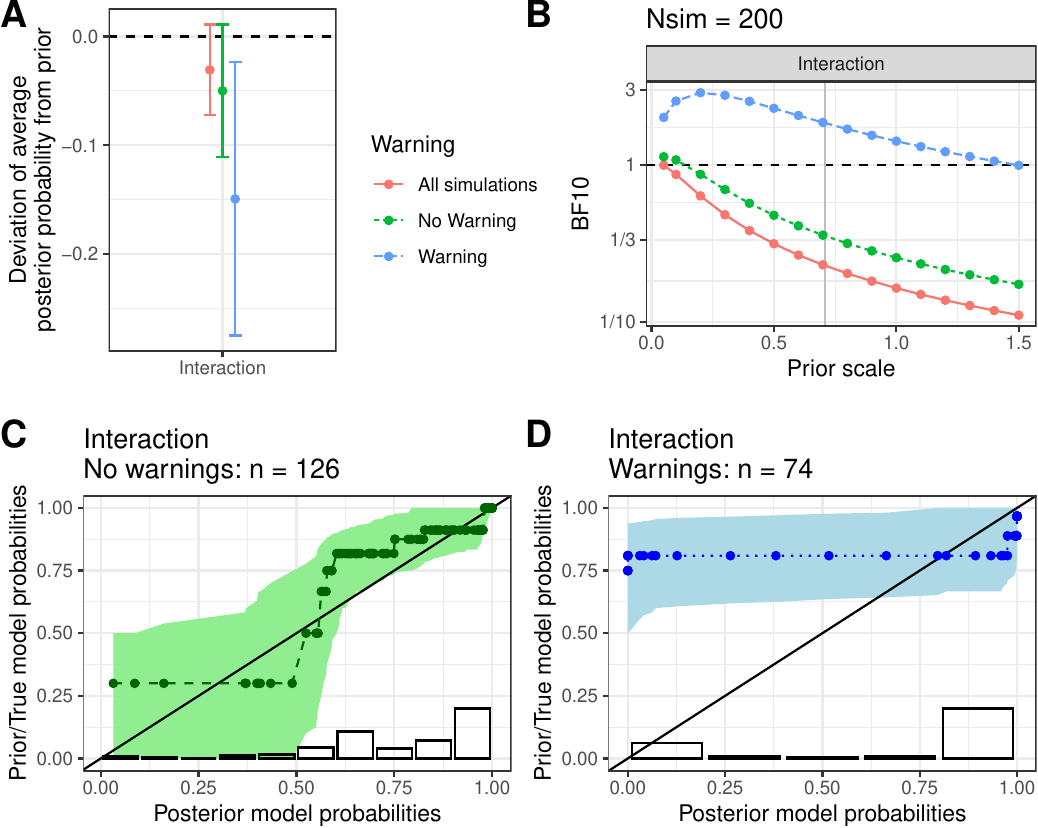} 

}

\caption{Repetition of the previous analyses of the interaction term  (in design 3), but with the prior probability for H0 set to 0.2 instead of 0.8 (10,000 iterations of MCMC sampling). A) The average deviation of posterior model probabilities for H1 from the priors (95\% CI). All simulations and simulations without / with a warning message. Deviations from the horizontal dashed line indicate bias. B) Bayesian t-tests of whether the posterior probability for H1 deviates from the prior, with sensitivity analyses for the prior scale. The vertical grey line indicates a default Bayes factor. C+D) Reliability diagrams: the prior model probability is plotted as a function of the estimated posterior model probabilities for simulations without (C) and with (D) a warning message. Error bands show 95\% consistency bands. Results show that when no warning is issued, the results are unclear. When a warning message is issued, clear bias is visible.}\label{fig:2stepItem-p080-plot}
\end{figure}

\textcolor{black}{We first investigated model convergence. Across all model fits} (\(200\) simulations \(\times 2\) models (H0/H1) \(= 400\) fits), \textcolor{black}{we observed} 0 \textcolor{black}{divergent transitions. R-hat exceeded the threshold of 1.01 for at least one population-level parameter in} 2 \textcolor{black}{cases, suggesting overall good model convergence.
However, during the bridge sampling, there was the warning message "logml could not be estimated within maxiter, rerunning with adjusted starting value. Estimate might be more variable than usual" in} 74 simulations (37.00\%)\textcolor{black}{, suggesting a similar amount of warning messages. In all} 74 \textcolor{black}{  cases, the warning message occurred for the H0 model. For the H1 model, there were warning messages in} 69 \textcolor{black}{cases. Out of the total $200$ simulations,} 31 \textcolor{black}{simulated the H0 and} 169 \textcolor{black}{simulated the H1. From these simulations, of the H0 simulations,} 10 (32\%) \textcolor{black}{exhibited a warning message, and of the H1 simulations,} 64 (38\%) \textcolor{black}{exhibited a warning message, suggesting similar percentages. We again looked at the results separately for simulations with versus without warning messages from the bridge sampler.}

\textcolor{black}{The results (see Fig.}~\ref{fig:2stepItem-p080-plot}\textcolor{black}{) showed that the average posterior model probabilities tended to be too small this time. Overall, across all simulations, there was a bias towards small posterior probabilities for the H1 (default} \(BF_{10} > 3\)\textcolor{black}{). The bias was numerically larger for simulations with a warning message (effect estimate), but the Bayes factor was undecided (presumably due to larger posterior uncertainty). For simulations without a warning message, the null hypothesis that the posterior model probability was equal to the prior was supported by the Bayesian t-tests and the sensitivity analysis (see Fig.}~\ref{fig:2stepItem-p080-plot}\textcolor{black}{B), showing some evidence for no bias (default} \(BF_{01} \approx 3\)).

\textcolor{black}{Reliability diagrams (see Fig.}~\ref{fig:2stepItem-p080-plot}\textcolor{black}{C+D) showed that simulations without warning message did not show clear evidence for bias. However, this was based on only few simulations (see histogram in Fig.}~\ref{fig:2stepItem-p080-plot}\textcolor{black}{C), and more SBC simulations may be needed to draw firm conclusions.
However, in simulations where a warning message was issued by the bridge sampler, the reliability diagrams suggested that posterior and prior model probabilities again did not match, indicating bias. Specifically, for low posterior probabilities (of H1) the prior probabilities were too high, now suggesting falsely low posterior model probabilities.}

\begin{figure}

{\centering \includegraphics{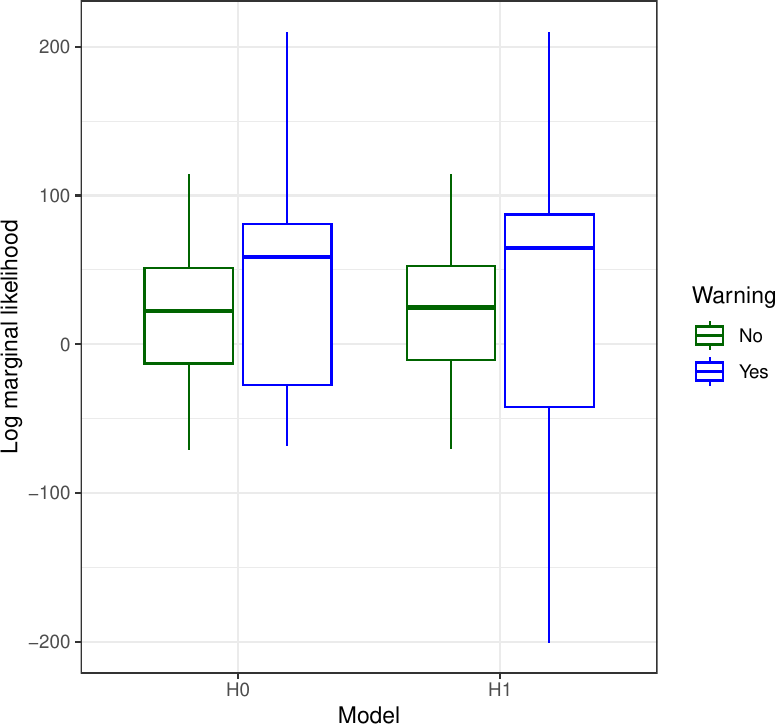} 

}

\caption{Repetition of the previous analyses of the interaction term  (in design 3), but with the prior probability for H0 set to 0.2 instead of 0.8 (10,000 iterations of MCMC sampling). Box plots of the log marginal likelihoods for models H1 versus H0 and for runs with versus without a warning message. Outliers removed in boxplots.}\label{fig:2stepItem-p080-ML-plot}
\end{figure}

\textcolor{black}{Again, we checked the distributions of the log marginal likelihoods for the H1 versus the H0 models across simulations. We found (see Fig.}~\ref{fig:2stepItem-p080-ML-plot}) \textcolor{black}{that the variances were again much enhanced when a warning message was issued by the bridge sampler.}

\textcolor{black}{The results therefore suggest that the observed bias for simulations with a warning message does not reflect a general upward-bias of the Bayes factors, but rather a noisy or uninformative result for the log marginal likelihood with relatively large variance, pushing the posterior model probabilities towards 0.5 on average due to floor and ceiling effects.}

\section{Discussion}\label{discussion}

In recent years, very user-friendly software has been developed to \textcolor{black}{approximate} Bayes factors in (generalized) linear mixed-effects models in the cognitive sciences, including the bridgesampling/brms packages. However, Bayes factors in complex models have to be \textcolor{black}{approximated}, with no guarantees for their accuracy. Here, we used marginal simulation-based calibration for Bayes factors (Schad et al., 2023) in combination with calibration plots (Modrák et al., 2025) to test the accuracy of Bayes factor estimation by studying some commonly used factorial experimental designs from the cognitive sciences. The results showed that when no warning message was issued by the bridge sampler, Bayes factors were estimated accurately in a range of well-studied repeated measures (Latin square) designs, which included (a; design 1) a \(2 \times 2\) design with random effects for subjects, (b; design 2) a design with a single fixed factor and crossed random effects for subjects and items, and (c; design 3) a \(2 \times 2\) design with crossed random effects for subjects and items. This provides important validation of the relevant Bayes factor estimates.
However, in simulations where a warning message was issued by the bridge sampler, we found clear biases in Bayes factor estimates indicating that estimated Bayes factors were \textcolor{black}{uninformative, of high variance, and did not correlate with the true prior model. In situations with a high prior probability for H0 (0.8), this resulted in Bayes factors that were} too large, and larger than the true Bayes factors, leading to liberal statistical null hypothesis tests that may indicate evidence for an effect even when the effect is in fact zero. \textcolor{black}{By contrast, in a situation with a high prior probability for H1 (0.8), this resulted in Bayes factors that were too small, and that tended to be smaller than the true Bayes factors, leading to conservative statistical null hypothesis tests that may indicate evidence for the null even when H1 is in fact true.}

These results are encouraging for the application of null hypothesis Bayes factor tests in common factorial designs, which are often used in fields such as psycholinguistics, psychology, and cognitive science.
At the same time, the results strongly suggest that warning messages of the bridge sampler need to be taken very seriously. When a warning message was issued by the bridge sampler, the resulting Bayes factor estimates were \textcolor{black}{highly variable,} biased and could not be trusted. Therefore, users of bridge sampling software should be aware that they should not report Bayes factor results when a warning message is shown, and that outcomes may not only be more variable than usual, but may also be (liberally \textcolor{black}{or conservatively}) biased.

One important question is what a researcher should do in case warning messages are issued by the bridge sampler. Often, those would correspond to computational problems and could be --- at least in principle --- resolved by a suitable reparametrization or other modification to the model, although this might be difficult in practice. Here, we also studied whether increasing the number of MCMC samples from \(10,000\) to \(40,000\) could reduce the number of warning messages. However, we found only a very small improvement. Note also, that we used the \texttt{method\ =\ "warp3"} argument of the bridge sampler, which may help to reduce warning messages, but warning messages remained. In case this should not be successful either, further options are to report biased Bayes factors together with a discussion of their bias, or \textcolor{black}{better} to turn to different methods for hypothesis testing. This could involve alternative methods for Bayes factor estimation (such as the Savage-Dickey method or the BayesFactor package) after testing whether they are accurate, but could also involve alternative approaches such as region of practical equivalence (ROPE) (Linde, Tendeiro, Selker, Wagenmakers, \& Ravenzwaaij, 2023), or posterior credibility intervals.

\subsubsection{Limitations}\label{limitations}

Overall, our results are limited in that we only focussed on a limited set of analysis settings, priors, models, experimental designs, and software packages. There is thus no guarantee that similar results will be obtained with slightly different designs, models, priors, and data sets. The present results therefore highlight that it is good practice to validate the use of Bayes factors when applying them in practice using simulation-based calibration for Bayes factors (Schad et al., 2023) or alternative approaches (Sekulovski et al., 2024). We hope that the present results will lead to a more robust use and application of Bayes factors by highlighting their strengths and limitations (flagged by warning messages) and also by encouraging researchers to check for their individual analysis, whether estimated Bayes factors are accurate.
However, performing SBC for every single analysis can be time consuming and difficult to implement for non-expert users. Therefore, we think that more future research is needed to provide a thorough characterization of the conditions under which Bayes factor estimates are biased versus accurate.

\section{Acknowledgements}\label{acknowledgements}

We thank Bruno Nicenboim \textcolor{black}{and Shravan Vasishth} for discussion.

\section{Declarations}\label{declarations}

\begin{itemize}
\tightlist
\item
  Funding
\end{itemize}

This work was partly funded by the Deutsche Forschungsgemeinschaft (DFG, German Research Foundation), Sonderforschungsbereich 1287, project number 317633480 (Limits of Variability in Language).

\begin{itemize}
\tightlist
\item
  Conflicts of interest/Competing interests
\end{itemize}

None

\begin{itemize}
\tightlist
\item
  Ethics approval
\end{itemize}

Not applicable.

\begin{itemize}
\tightlist
\item
  Consent to participate
\end{itemize}

Not applicable.

\begin{itemize}
\tightlist
\item
  Consent for publication
\end{itemize}

Not applicable.

\begin{itemize}
\tightlist
\item
  Availability of data and materials
\end{itemize}

Not applicable.

\begin{itemize}
\tightlist
\item
  Code availability
\end{itemize}

The code is available on OSF (\url{https://osf.io/3g86r/}).

\begin{itemize}
\tightlist
\item
  Authors' contributions
\end{itemize}

DJS: Conception, methodology, analysis, writing of first draft, review and editing.

MM: Methodology, analysis, review and editing.

\section{Open Practices Statement}\label{open-practices-statement}

Analysis code is available at \url{https://osf.io/3g86r/}. The reported studies were not preregistered.

\newpage

\section{References}\label{references}

\begingroup
\setlength{\parindent}{-0.5in}
\setlength{\leftskip}{0.5in}

\protect\phantomsection\label{refs}

\endgroup

\end{document}